\documentclass[12pt]{report}
\usepackage[dvips]{graphics}
\usepackage{amsmath}
\usepackage{xspace}
\usepackage{epsfig}

\newcommand{\numub}{\ensuremath{\overline{\nu}_\mu}\xspace}
\newcommand{\nueb}{\ensuremath{\overline{\nu}_e}\xspace}
\newcommand{\numu}{\ensuremath{\nu_\mu}\xspace}
\newcommand{\nue}{\ensuremath{\nu_e}\xspace}  
\newcommand{\nux}{\ensuremath{\nu_x}\xspace}  

\newcommand{\nutau}{\ensuremath{\nu_\tau}\xspace}

\newcommand{\dmq}{\ensuremath{\Delta M^{2}}\xspace}

\begin{document}
\LARGE{Bruno Pontecorvo and neutrino physics}
\vskip 1cm
\par
\Large{Ubaldo Dore \\
Dipartimento di Fisica, Universit\`a di Roma ``La Sapienza",\\
and I.N.F.N., Sezione di Roma, P. A. Moro 2, Roma, Italy\\}
\Large{Lucia Zanello \\
Dipartimento di Fisica, Universit\`a di Roma ``La Sapienza",\\
and I.N.F.N., Sezione di Roma, P. A. Moro 2, Roma, Italy\\}


\begin{abstract}
In this paper  the contribution of  Bruno Pontecorvo
in the field of neutrino physics will be illustrated. Special emphasis 
will be given
to the  physics of oscillations that he was the first to propose .    
\end{abstract}
\newpage

\tableofcontents
\newpage
\section{Introduction}
\par This paper is dedicated to illustrate
the Bruno Pontecorvo contribution to neutrino Physics.
\par BP contribution to neutrino physics make him one the major
contributors to the development of particle physics in the XX century.
His ideas and proposals were then the subject of many successful
experiments that gave to their authors Nobel Prize  award. We will describe
his ideas in this paper.
Many ideas about neutrino physics were put forward by BP
  much in advance of the times.  The most
important  were:
\begin{itemize}
\item   As soon as nuclear reactors
were built he realized the fact that the produced intense flux of neutrinos
would  have  made the detection of these particles possible at a time in
which many physicists thought  that this observation was impossible.
\item He suggested the use of the neutrino chlorine reaction that was then
used
in the Homestake experiment that started the experimental oscillation physics.
\item The results of the Conversi-Pancini-Piccioni experiment brought him
to a first
introduction of the concept of Universality in weak interactions.
\item Neutrino experiments at accelerators and the existence of two types
of neutrino was anticipated by BP.
\item The hypothesis of neutrino mixing and so of oscillations was put
forward by BP.
 The existence of oscillations and so the fact
that neutrinos have mass has been
 one of the most important results of  particle physics in the
last years.
\end{itemize}
\par The paper is  organized in the following way
\begin{itemize}
\item Brief introduction to neutrino physics
\item Biography of Bruno Pontecorvo
\item The scientific production
\item The big intuition "OSCILLATIONS".
\end {itemize}
\begin{figure}[h]
\begin{center}
\epsfig{figure=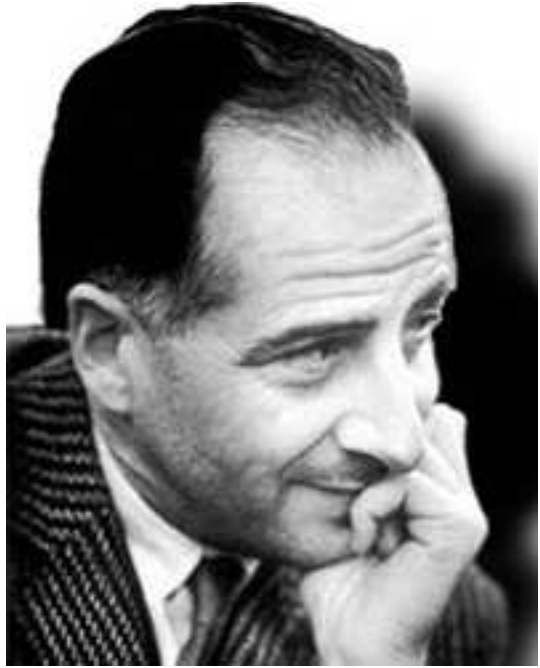}
\label{pontecorvo}
\caption {Bruno Pontecorvo }
\end{center}
\end{figure}

\section{History of neutrino}
\subsection{The birth of neutrino}
\par In 1930 Pauli \cite{pauli} suggested, to save energy conservation in 
beta decay,
that a  neutral light particle was  emitted in the process, he named it 
{\it neutron} in
June 1931.\par  In 1932 the neutron was discovered by Chadwick \cite{cha} so the
new Pauli particle
was called {\it neutrino }by E. Fermi.
\subsection{The detection of neutrino}
\par The smallness of  cross section made the neutrinos still a
hypothesis; many
physicists thought  that neutrino detection was almost impossible.
\par Pontecorvo  showed \cite{inverse}  (see paragraph \ref{det}) that
 in fact  the
detection could be possible.

\par The detection of free neutrinos was accomplished by Cowan and Reines
\cite{reines}
 who observed the reaction
 $$ \nueb +p \rightarrow n+e^+$$
   This experiment was the first  to detect free neutrinos and Reines
was  awarded  the Nobel prize in 1995.
\subsection{Present knowledge of neutrino physics}
\par For a complete review of current status of knowledge on neutrino 
physics see \cite{PDBook} 
\par We summarize briefly our present knowledge of neutrino properties:

\begin{itemize}
\item Neutrinos are chargeless fermions interacting only through weak
interactions.
\item Neutrinos interact through the exchange of  W (charged currents) or
Z0 (neutral currents).
\item the V-A theory requires, in the limit of massless neutrinos, that
only left handed neutrinos be active. The opposite   is true for
antineutrinos.
\item In the Minimum Standard Model(MSM) there are 3 type of massless
neutrinos (we shall see that the massless  condition must be relaxed) and
a
corresponding number of antineutrinos.
\item Weak interactions have the same strength for the three species,
(Universality).
\item Neutrinos are coupled to the corresponding charged leptons so we
have 3 leptonic doublets
\par
 $\begin{pmatrix}e^{-}\cr\nue\end{pmatrix}$,
 $\begin{pmatrix}\mu^{-}\cr\numu\end{pmatrix}$,
 $\begin{pmatrix}\tau^{-}\cr\nutau\end{pmatrix}$.
\item Leptons  in each doublet carry an additive leptonic
number $L_e$, $L_\mu$, $L_\tau$, which has opposite
sign for antiparticles.
\item Leptonic numbers are separately conserved.
\end{itemize}
\par There are  still  questions that must be answered.
\par Are neutrinos Dirac or Maiorana particles?
\par Neutrinos have a mass but the absolute scale of this  mass
is still unknown

 \section{Biography}
Bruno Pontecorvo was born in Pisa in 1913 from a known
jewish family. He had three   brothers and two  sisters , one brother 
was
Guido(1907-1999) professor of genetics, the other was Gillo(1919-2008) the 
well-known  film director.
\par In 1929 he left Pisa and
enrolled in the third year of the degree in Physics  at the  University 
of Rome.
After  his graduation he became the youngest member of the Fermi Group. 
In 1934 he collaborated to   the  famous experiment on slow neutrons
\cite{fermi1},\cite{fermi2}.

In 1936 he moved to Paris in the Joliot Curie laboratory. He was of a
jewish  family, so after the German invasion he had to flee to Spain
 and then to the United States with his family:
his wife  Marianne Nordblon and his first son Gil.
He found  a job in an  oil search company. In 1943 he was called
to participate to the construction of a nuclear reactor in Canada. He
stayed there until 1948. In that period his sons  Tito and Bruno were  
born.In 1948 he became a British citizen  and moved
to England called by J. Cockroft.
In 1950 he traveled  to  Italy, officially on holiday. He went with his 
family to
Stockolm and then to  the Soviet Union in the Dubna laboratories.
\par In Soviet union he was called Bruno Maximovic.
\par From Dubna he made several trips to Italy starting in 1978.
\begin{figure}[h]
\begin{center}
\epsfig{figure=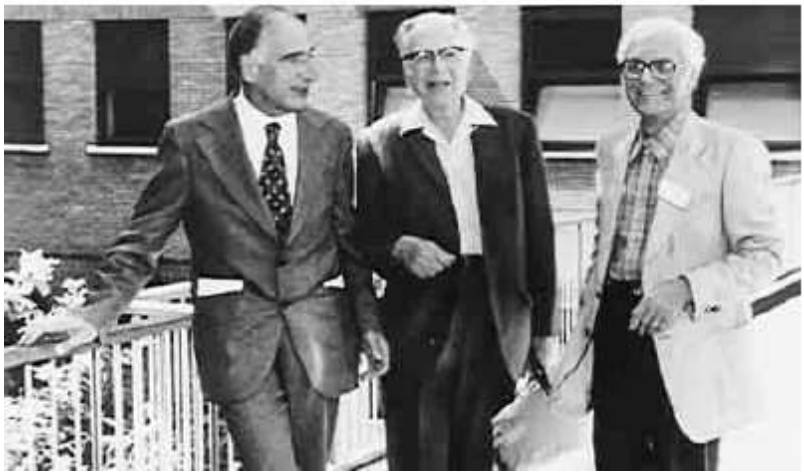}
\label{pontecorvo}
\caption {Bruno Pontecorvo,Emilio  Segre and Edoardo Amaldi
in Rome 1978 }
\end{center}
\end{figure}
\par The last ten years of BP life were a courageus struggle against 
Parkinson disease.  He never stopped to work  on physics and oscillations
\begin{figure}[h]
\begin{center}   
\epsfig{figure=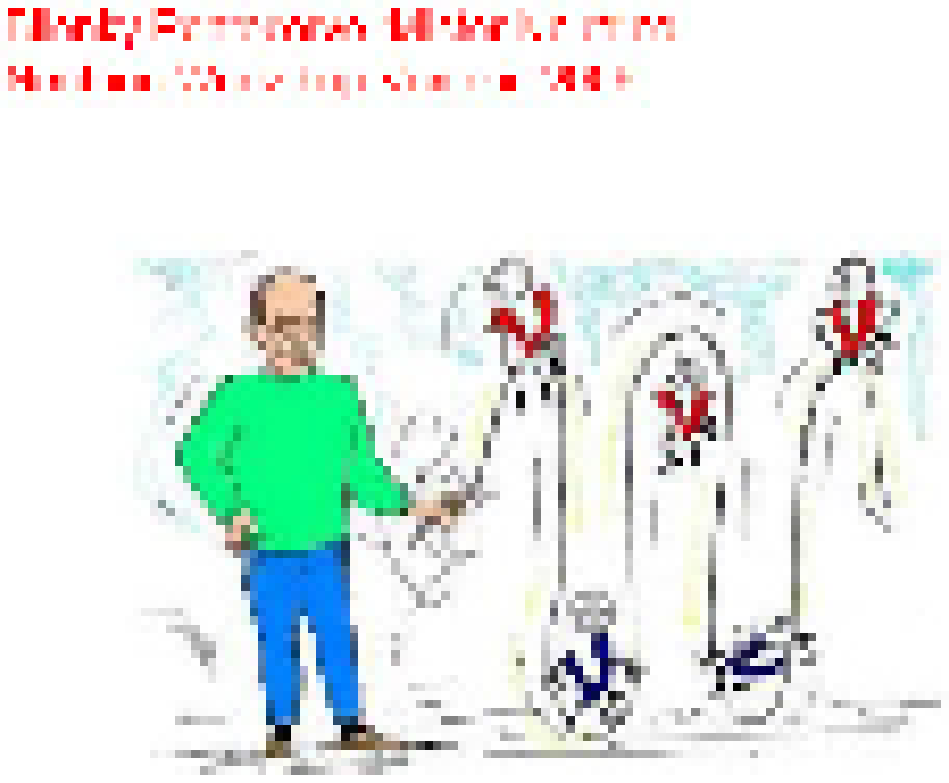}
\label{pontecorvo1}
\caption {Dubna 1988 Neutrino oscillations}
\end{center}
\end{figure} 
\begin{figure}[h]
\begin{center}\epsfig{figure=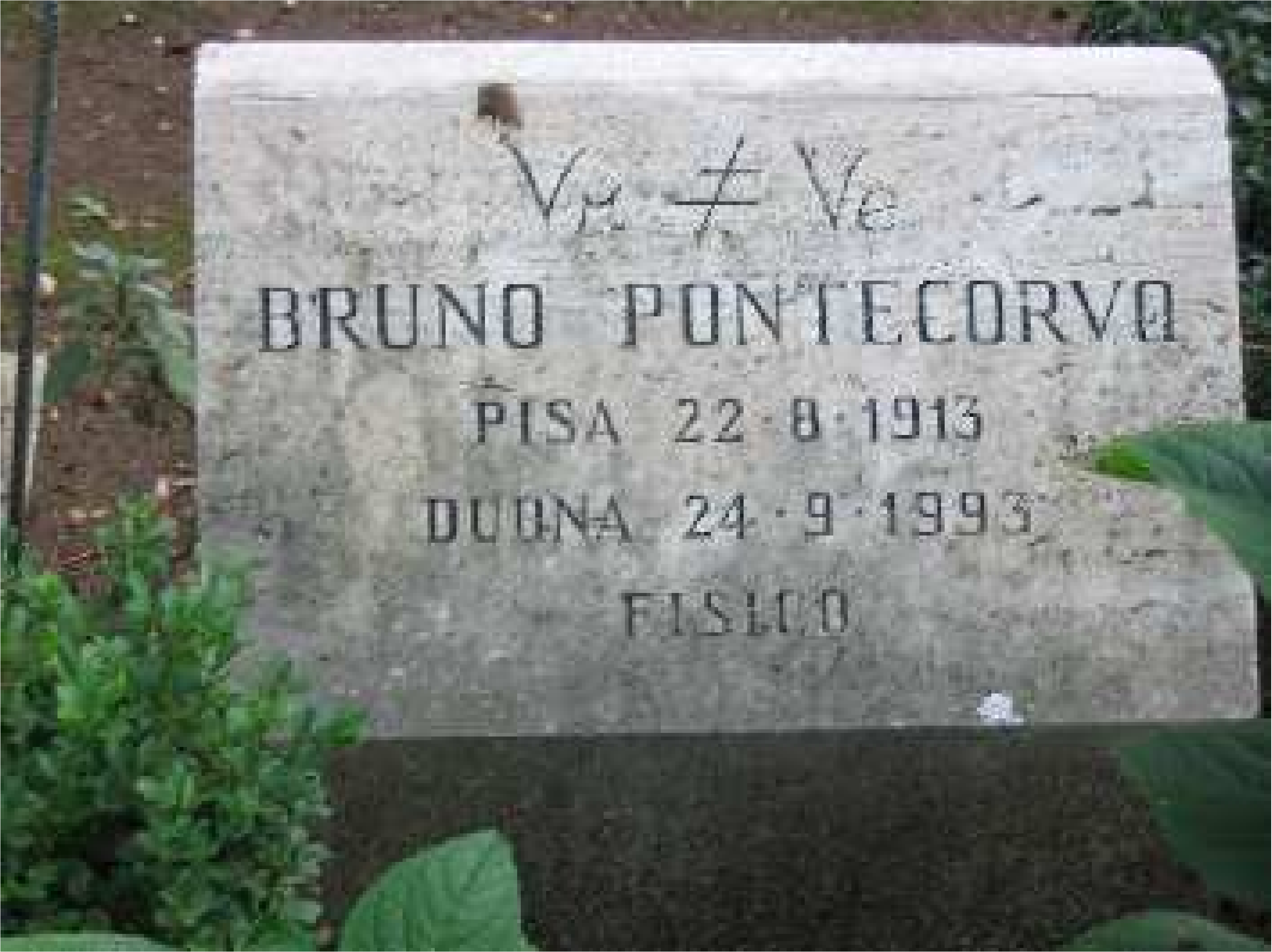,width=12.cm}
\label{pontecorvo3}
\caption { Bruno Pontecorvo tombstone. foto Alvaro De Alvaris 
http://www.flickr.com/photos/dealvariis/ }
\label{tomba}
\end{center}  
\end{figure}

 He remained in the Soviet Union until his death in 1993.
\par Now his tombstone is in the ´Cimitero Acattolico´ in Rome. 
the tomstone is shown in figure \ref{tomba} 
\par For the  reasons that made him  decide to go to the  Soviet Union
one can quote  the words of V.P. Dzhelepov \cite{dzel}, a distinguished
russian physicist:
\par {\sl Bruno was an Italian communist, at the time of arrival in the
Soviet Union
he was a communist-idealist sincerely believing in the strength and
rightness
of the type of  development chosen by Russia.}
\par It must be noted that there was a malevolent interpretation of his 
arrival in USSR :he was a soviet spy and he fled before being unmasked.

\par the Impact of BP on neutrino physics is well 
recognized in the Scientific world.
\par we will give  three  examples 
\par  Valentine Telegdi quoted by L.Okun \cite{okun} 
\par {\sl Almost all important idea in neutrino physics are due to 
Pontecorvo}
\par Jack Steinberger \cite{steinb}
\par{\sl Few of us in particle physics  can boast of a single original 
and 
important idea. Bruno wealth of seminal suggestions establish him as a 
truly unique contributor to the the advance of particle physics in the 
past 
half century}
\par Nicola Cabibbo \cite{cabi}

\par {\sl Con Pontecorvo scompare uno dei grandi scienziati del XX 
secolo 
uno dei pionieri della fisica delle alte energie.}

\section{Scientific activity}
\par BP  started his scientific activity in Rome in the group of E.
Fermi,
participating to  the famous experiments on the slowing down of neutrons
in
hydrogenous materials \cite{fermi1},\cite{fermi2}
\par From 1936 to 1938 he worked  in Paris with F. Joliot Curie on
nuclear isomerism \cite{isomer}
\par From 1938 to 1940 he worked in the 'Well survey inc', Tulsa USA
where he developed a system \cite{logging} for finding oil or water
underground.
The system  used a neutron source and was the first application of the
Rome results on the slowing down of neutrons.
In the following, activities in Canada and Dubna will be presented.

\subsection{Activities in  Canada }
\par BP  lived in Canada from 1943 to 1950. At the beginning he stayed
in Montreal
where he worked on the NRX project, a heavy water, natural uranium
 reactor.
He  worked  on the design of the reactor that started in 1947 yielding a
flux
of  6.10$^{15}$ \nueb/cm$^2$ sec

 The reactor was
then built near the  Chalk river , and he lived in the nearby Deep
River town.
We want to recall his contributions on the following arguments
\begin{itemize}
\item{muon decay}
\item{universality}
\item{neutrino detectors}
\item {proportionals counters}
\end{itemize}
\par \underline{muon decay}
\par In 1950 he published, together with E.P.Hincks  \cite{muon}, the
result of
an
experiment on the decay of the penetrating cosmic ray particle, the muon,
that was known
to have a 2.2 $\mu$s lifetime, while    decay products were not
known.  The only thing known was that  one charged
particle was emitted but there was no information on its nature and on the
nature of the  emitted neutral particle(s).
In the experiment, muons were stopped in an absorber and were  detected
by an arrangement of Geiger-Muller counters.
The decay products were  detected with  a delayed coincidence technique.
The results  of the experiment were
\begin{itemize}
\item the charged particle emitted is an electron
\item the average energy of the electron is greater than 25 Mev
\item there is no evidence of emission of electrons with energy
       larger than 50 MeV.
\item the  shape of the energy spectrum of the electron excludes the possibility
      of being a single line
\end{itemize}
\par The conclusion was
\par {\sl The average energy and the form of the energy spectrum
of decay electrons
 are, within the accuracy of theory and experiments, in agreement
 with theoretical expectations of a process}
 $$\mu \rightarrow e + \nu +\nu$$

\par\underline{Universality}
\par The similarity of e-N and $\mu$-N interactions later known as
Universality was for the first time pointed out by BP.
\par In 1946 a paper of Conversi, Pancini and Piccioni \cite{conversi} was
published on the results of  the behaviour of the
negative and positive particles of the hard component of cosmic rays.
This experiment   showed that negative
particles where not captured when stopped in light elements as was
expected if the negative particle had  been the Yukava particle.
 Lattes, Occhialini and Powell \cite{occhia}  in  1947 in fact  showed
that
cosmic ray
muons were the decay products of a particle now known as $\pi$.
Thinking about the Conversi experiment, BP made some relevant
considerations
as shown in the letter he wrote to Giancarlo Wick in
1947.
\par letter to  GC Wick \cite{wick}

\par {\sl Deep River 8 maggio 1947,

  Caro Giancarlo ... se ne deduce una similarita' 
tra processi beta e processi di assorbimento
ed emissione di mesoni, che, assumendo non si tratti di una coincidenza,
sembra di carattere fondamentale.}
\par English translation
\par {\sl It can be deduced a similarity between beta processes and 
processes 
of 
absorpion or emission of mesons, that, assuming that it is not 
coincidence, seems to be of fundamental character}

\par He then published his considerations \cite{univ}

\par
{\sl We notice that the probability (10$^8$ sec$^{-1}$) of capture
of a bound
negative meson is of the order of the probability of ordinary K-capture
processes, when allowance is made for the difference in the
disintegration energy and the difference in the volumes of the K-shell
and of the meson orbit.
We assume that this is significant and wish to discuss the possibility of
a fundamental analogy between beta-processes and processes of emission and
absorption of charged mesons}
\par This paper introducing the concept  of Universality had not a large echo.
 In fact after 2 years there was a paper
of G.Puppi that introduced the concept of a new fundamental force,
the 'Universal weak interaction',
\cite{puppi} in which there was no reference  to the Pontecorvo ideas. The
same  happened in the papers of other authors all concerning the
universality of the weak interactions \cite{klein},
\cite{lee},\cite{tiomno}

\underline{Neutrino detectors} \label{det}
\par In 1946 in his paper 'inverse  beta processes ' \cite{inverse} BP
discussed the possibility of detection of
neutrinos. Although the detection of the inverse beta decay
           $$\nu +Z \rightarrow (Z-1) +e^{+} $$ was at that time
considered
not possible. due the
calculated cross section $10^{-44} cm^{2}$ given
by Bethe and Peierls \cite{bethe}, the use of powerful neutrino sources as the nuclear
reactors could make the detection of the above process possible.
He wrote

\par{\sl It is true that the actual $\beta$ transition is certainly
 not detectable in practice. However
the
nucleus  of charge (Z$\pm$1) produced in the reaction
may be radioactive with a decay period well known. The
essential point in this method is that the radioactive atoms produced
must have  different chemical
properties of the irradiated atoms. Consequently it may be possible
 to concentrate the radioactive atoms from
a very large volume}
\par The principal requirements for the irradiated material had to be
\begin{itemize}
\item the irradiated material must be not too expensive
\item the produced nucleus must be radioactive with a period of at
least one day, because of the long time
involved in the separation
\item The separation of radioactive atoms must be relatively simple
\item the background must be as small as possible
\end{itemize}
\par BP suggested the reaction
 $$\nu +^{37}Cl \rightarrow ^{37}Ar +e$$
 As a source of neutrinos he suggested
\par a) neutrino flux from the sun
\par b) neutrinos from recently developed nuclear reactors
\par It must be noted that the sun produces neutrinos, while in reactors
antineutrinos
 are produced,
at that the difference between neutrinos and antineutrinos was not yet clear.
\par The above process was used in the solar neutrino Homestake Davis
 experiment that started in 1962. The neutrino deficit
compared with the
prediction of the standard  solar model gave  origin to the "neutrino puzzle'.
 The final result of the
experiment, that lasted several decades, were published in 1998
\cite{davis}.
In this paper the
Pontecorvo suggestion was recognized.
\par In 1946 the detection of the inverse beta decay process
          $$\overline\nu +p \rightarrow n +e $$
with neutrinos from nuclear plants
was not feasible.
\par The development of the liquid scintillator technique ten  years later
 allowed
 the detection
of the above process.\cite{reines}. The Reines experiment was  the first
experimental proof of the existence of the neutrino.

\underline{proportional counters}
\par BP together with Hanna and Kirkwood developed  a new technique of proportional counters,
based on very large amplification in the gas. Sensitivity to a few ion
couples
 was reached \cite{hanna}.
This development was essential in the solar neutrino Davis experiment
 and later in the He$^3$ proportional
counters of the SNO experiment.
\par The new technique  was used in experiments of low energy
spectrometry and used  in the measurement
of the  tritium $\beta$ spectrum.

In the experiment, published in 1949, a first measurement of the neutrino
mass
was obtained \cite{massa}.The obtained value was
m$_\nu \leq 500~ eV$

\subsection{Activities in Dubna}
\par In
1950, BP  started his activity  in the Dubna laboratories (JINR).
His notes and reports were in Russian. The
English
translation   of part of this material
can be found in ´Selected Scientific Works recollection
on Bruno Pontecorvo´ \cite{sif}.
\par He  participated in experiments at the Dubna and Serpukov
accelerators.
\par Out of the  experiments at the Dubna Synchrocyclotron  we can recall
\begin{itemize}

 \item muon capture  \cite{capture}
\item  measurement of the pion interactions\cite{pion}
\item search for the production of $\Lambda$ in proton interactions at 700 MeV
\cite{lambda}.
\item search for anomalous scattering of muon neutrinos by nucleons
\cite{anomal}
\end{itemize}
 Results of experiments at the Serpukhov accelerator can be
found in \cite{serpu},\cite{serpu1}.
\par There are several arguments that he considered. We will give a brief
resume.
(One of the  more important contribution of Pontecorvo, Oscillations, will
be considered
 in next section \ref{oscil2})
\par \underline{Neutrino beams}
\par In  1959 BP  started to think to  the possibility of performing experiments
 with neutrino
produced  at accelerators. The first problem to be solved was the
possibility of the existence  of two types of neutrinos, namely
neutrinos from beta decay and neutrinos from pion and muon decay.He
discussed the  problem in \cite{two}. One year later M. Schwarz discussed
the same problem \cite{schwa}.  Schwarz then partecipated in an experiment at
the AGS in Brookhaven USA  together with L. Lederman and J. Steinberger.
They proved  that there are two type of neutrinos \cite{bnl} named $\nue$
and $\numu$.For this result
the three authors were awarded  the Nobel prize in 1988.
BP also proposed a non conventional technique to produce neutrino beams.
\par Conventional high energy neutrino beams are produced by pion and K
decays
produced in proton interactions . Their flavor content is:  neutrinos
,(antineutrinos) coming from the decay
of ($\pi,K)^{+}$and  ($\pi,K)^{-}$. These  beams have a small $\nu_e$
contribution coming from the three  body decay of K mesons.
Bruno's  innovative idea was   the "beam dump" \cite{dump},
a technique proposed to search for short lived particles
 that decay before interacting. Protons are made to interact  in heavy
materials.
Pion and Kaons  interact before decaying, so the only produced neutrinos
 come from the decay of short lifetime particles (charms).
In this case the contribution of $\nue$ and $\numu$ are comparable.
\par \underline{Neutrino and astrophysics}
\par Pontecorvo  published  several papers on this argument.
Many of these are given in the following table
\begin{table}[htbp]
\begin{center}
\begin{tabular}{|c|c|c|c|c|}\\\hline
 n &{\bf year} & title &refer \\\hline
 1&1959 & Universal Fermi interactions and astrophysics
&\cite{astro1}\\\hline
2& 1961&Neutrino and  density of matter in the Universe
&\cite{astro2}\\\hline
3&1963&Neutrino and its role in Astrophysics&\cite{astro3}\\\hline
4&1969&Neutrino Astronomy and lepton charge& \cite{astro4}\\\hline

\end{tabular}
\end{center}
\end{table}
\par
All these papers can be found translated in English  in ref \cite{sif}

\section {Oscillations} \label{oscil2}
\par The contribution of Bruno Pontecorvo to the field of neutrino
 oscillation was fundamental and he
defended the concept of oscillations in years in which
the neutrino were considered massless and so
oscillation impossible. In this section we will present
\begin{itemize}
\item {Basics  of neutrino oscillation theory}
\item {Experimental results}
\item {The Bruno Pontecorvo contribution}
\end {itemize}
\subsection {Basics  of neutrino oscillation theory}
\par As in the quark  sector  the weak interactions states are  a linear
superposition of the mass eigenstates.
These states are connected by an unitary matrix U
$$ \nu_{\alpha}=\sum_j U_{\alpha j}\cdot\nu_j $$
with index $\alpha$ running over the three flavor eigenstates and index j
running over the three mass eigenstates.
\par The matrix U is called the Pontecorvo--Maki--Nakagawa--Sakata.
 In the general case, a $3\times3$ matrix can be parametrized by 3 mixing
angles $\theta_1=\theta_{12}$, $\theta_2=\theta_{23}$, $\theta_3=\theta_{13}$
and a CP violating phase $\delta$.
A frequently used parametrization of the U matrix is the following
\begin{eqnarray}\nonumber
U&=&\begin{pmatrix}1&0&0\cr 0&c_{23}&s_{23}\cr 0&-s_{23}&c_{23}
\end{pmatrix}
\begin{pmatrix}c_{13}&0&s_{13}e^{-i\delta}&
 \cr
0&1&0\cr -s_{13}e^{+i\delta}
 &0&c_{13}
\end{pmatrix}
\begin{pmatrix}c_{12}&s_{12}&0\cr -s_{12}&c_{12}&0\cr 0&0&1

\end{pmatrix}
\end{eqnarray}
\par where $c_{jk}=\cos(\theta_{jk})$ ~~~and ~~~~  $s_{jk}=\sin(\theta_{jk})$.
\par  Given three neutrino
masses, we can define two independent
 square mass differences $\dmq_{12}$ and  $\dmq_{23}$.
 $\dmq_{12} =M_{1}^2- M_{2}^2$,
 $\dmq_{23} =M_{2}^2- M_{3}^2$

\par It has  experimentally been shown that
$|\dmq_{12}|\ll |\dmq_{23}|$ and so $\dmq_{13}\simeq
\dmq_{23}$.
\par The mass spectrum is formed by a  closely spaced doublet  $\nu_1$ and
$\nu_2$,
 and by a third state $\nu_3$ relatively distant.
\par In many cases oscillations can be studied considering the simplified
approximation of  two  family
mixing. With two mass states  $M_1$ and $M_2$, the mixing matrix is
reduced to $2\times2$ and is
characterized by a single parameter,the mixing angle $\theta$ (omitting
irrelevant phase factors):

\begin{gather*}
\begin{pmatrix} \cos\theta&\sin\theta\\-\sin\theta&\cos\theta
\end{pmatrix}
\quad
\end{gather*}
\par Consider for example a $\nue$ beam and $\nue ~ \rightarrow \numu$
oscillations
\par At a distance L from
from the source the probability of detecting a $\nue$ as
$\numu$, is
\par $$ P(\nue \rightarrow \numu)
=\sin^{2}(2\theta) \sin^{2}(\Delta
M^{2}L/4E)$$
\par where $\dmq= M_1^{2}-M_2^{2}$.
~Choosing to express $\dmq$ in $eV^2$, L in m(Km) E in MeV(Gev) we have
\par $$ P(\nue \rightarrow \numu)=
\sin^{2}(2\theta) \sin^{2}(1.27 \Delta M^{2}L/E)$$
\par we emphasize that oscillations are sensitive only to the difference
of the square masses.

\subsection  {Experimental results . Solar neutrinos }
\par \underline{Radiochemical experiments}

\par The experimental story of the oscillations started with the solar
 neutrino
'puzzle'.
 The flux of neutrinos coming from the sun, as detected  in the Homestake
neutrino detector, was below the expected one.
Nuclear reactions in the Sun giving rise to neutrinos start with the PP
reaction
$$ p+p \rightarrow H^{2}+e^{+}+\nue$$
neutrinos produced in this reaction constitute 99$\%$ of all neutrinos emitted
 by the sun .
The energy spectrum of
neutrinos has an end point at  0.42 MeV.Additional  neutrinos are emitted
 in the chain
 process
initiated by
thr PP  reaction
 \par $ H^2+p\rightarrow He^3+\gamma$
\par
$He^3+He^4\rightarrow Be^7+\gamma$
\par $Be^7+e^-\rightarrow Li^7+\nue $
\par $Be^7+p\rightarrow B^8+\gamma$
\par $(Be^8)^*\rightarrow 2He^4$
$B^8\rightarrow (Be^8)^*+e^++\nue$
\par The net result of the chain is
$$ 4p +2 e^{-} \rightarrow He^{4}+2 \nue +\gamma.$$
The Q of the reaction is 26 MeV and the neutrinos take away on the average
0.5 MeV.
\par In the Davis experiment
solar netrinos coming from
 the sun were detected
via the reaction $$\nue +Cl^{37} \rightarrow Ar^{37} +e$$
as suggested  by BP.
\par The Davis detector was  a large tank containing 100000 gallons of
 tetrachloroetilene. 	It was located
in the Homestake mine at a depth of 4800 meters.
\par  Using physical and
chemical methods the amount of produced $Ar^{37}$ was extracted from the target
material. The  $Ar^{37}$  is unstable, so the extraction had to be performed
 periodically
Auger electrons or photons emitted in the decay were detected in proportional
 counters.

The experiment  ran from 1970  to 1995  and the  results were compared
with the model of neutrino emission from the sun the ¨Standard Solar Model¨ (SSM)
The main contributor to  the computation of SSM model and of his results 
was
J.Bahcall \cite{bahcall}.
The final result of the experiment was
$$\Phi(Davis)/\Phi(SSM)=0.34\pm 0.04$$
\par Although final results were published in 1998 \cite{davis},
preliminary were
published before , see for example \cite{bahcall2}.
\par The disagreement between SSM  and results, the 'solar neutrino
puzzle' was largely discussed in the physics community and many
explanations were given, but only
BP indicated oscillations as the origin of the  discrepancy.

\par Following the Davis results other experiments were done.

\par  The Gran Sasso Laboratories (Italy)experiment Gallex
\cite{Gallex} and then
GNO \cite{gno}and Baksan(Russia)\cite {sage}
radiochemical  experiments
studied  the process
$$ \nue +Ga^{71} \rightarrow Ge^{71}+e^{-} $$
The threshold of this reaction is 0.223 Mev so it is sensitive to the
PP reaction that has an energy endpoint at 0.42 Mev while the Chlorine experiment
had a threshold of 0.813 Mev and therefore  was not
sensitive to
the PP   reaction that is the origin of the the large majority of solar
neutrinos.
\par The weighted average of all Gallium results is~\cite{gavrin}
$$ Capture~ Rate =67.6\pm 3.71~~ SNU$$
to be compared with the prediction of 128 of the SSM
(1 SNU(standard neutrino unit)=$10^{-36}$neutrino captures/(atom sec)).
\par \underline{Real time experiments}
\par In the Kamioka mine in Japan the experiment Kamiokande ran from
1983  until 1988 followed by the
 Superkamiokande one, that  started in 1996, and  is still running.

\par Both experiments were large water Cherenkov counters (Kamiokande
 3  ktons, Superkamiokande 50 ktons)

in which the Cherenkov light emitted by fast particles  is  collected by
photomultipliers.
\par Solar neutrinos were detected via the reaction
  $$\nux +e \rightarrow \nux +e$$
\par all neutrinos contribute to the reaction but the main contribution is
given by $\nue$ because in this case we can have both charged current (CC)
interaction and neutral current(NC) ones, while in the case of $\numu$ and
$\nutau$ only neutral current reactions are allowed. The ratio is
CC/NC=$\sim$ 6

\par Results of the experiment still confirming the neutrino deficit
 are \par
Kamiokande \cite{KA}
flux
$$2.80  \pm 0.019(stat.) \pm 0.33(sys) .  10^6 cm^{-2} sec^{-1}$$
ratio
$\Phi(\nux)/\Phi(SSM)=0.55 \pm 0.19 \pm 0.33$
\par Superkamiokande \cite{SK1} flux
$$2.35  \pm 0.02(stat.) \pm 0.08(sys) . 10^6 cm^{-2} sec^{-1}$$
ratio
$\Phi(\nux)/\Phi(SSM)=0.47 \pm 0.04 \pm 0.14$
\par The Superkamiokande  result refer to phase 1 of the experinent. New results
have been published \cite{SK2} confirming the above  results.
\par \underline{Confirmation of the oscillation hypothesis}
\par The results on the ratio given above  rely on the correctness of
the SSM
and so the interpretation of the value of the ratio
 as due to oscillations   needs confirmation.

\par The confirmation  came in the year 2003 'annus mirabilis' from two
experiments,  SNO and Kamland.
\par The SNO experiment \cite{SNO}, Sudbury Neutrino observatory, was
a 1000 tons heavy water Cherenkov detector.
In Deuterium the following three reactions were  observed:
\par 1)  $\nue+d\rightarrow p+p+e^-$ charged current interaction
accessible only
to $\nue$
\par 2)$\nu_{x}+d \rightarrow p+n+\nu_{x}$   neutral current interaction
accessible to all neutrinos.
This is an unique feature of the SNO experiment that allows a direct verification of the SSM.
\par 3) $\nu_{x}+e\rightarrow \nu_{x}+e$ accessible to $\nue$ and, with
smaller cross section, to $\numu$ and $\nutau$.
\par Reactions 1 and 3 were observed via the detection of the Cherenkov
light emitted by electrons, while   reaction 2 was detected via the
observation of
the neutron in the final state.Various neutron detection techniques were
used in successive
parts of the experiment.
The results were summarized in the following way
$$R_{ee}=\Phi(CC)/\Phi(NC)=\Phi(\nue/\Phi(total)=0.34
\pm0.023^{+0.029}_{-0.031}$$
\par The $\nue$ flux is depressed, in agreement with the results of all
solar experiments
$$\Phi(\nux)/\Phi(SSM)=1 \pm 0.1$$
\par this is the fundamental result of the experiment.
\par The flux of all types of neutrinos is in good agreement with the SSM
 which is therefore confirmed.
\par Because the $\nue$ are depressed this means that neutrinos in their
path from Sun to Earth  have changed their flavor.
\par The Kamland experiment
\par The detector is located  in the Kamioka mine in Japan. It consists of
1 kton scintillator contained  in a balloon viewed by photomultipliers.
53 nuclear reactors surround Kamland at an average distance of 150 km.
Emitted
anti neutrinos interact in the hydrogen of the scintillator
 $$ \numub +p \rightarrow n+e^+$$
 The reaction products are  detected as a
pulse delayed
pair, the first pulse being due to the annihilation of the positron, the
second
to delayed gammas emitted in the  capture of the moderated neutron.
This is the same technique used in the first detection of free neutrinos
and in all $\nueb$ reactor neutrinos detection experiments.
The survival probability, ie the probability that antineutrinos originated
in the reactors
reach the detectors  has been
computed to be
$$ 0.658 \pm 0.044 stat \pm 0.047 syst$$
\par  Interpreting this result in
terms
of oscillations one obtains the same parameters that have been found in the
analysis of solar neutrinos.
\par  A global two flavor analysis of Kamland  data and solar
data~\cite{kamland}
 gives $$\dmq=(7.9^{+0.6}_{-0.5})\times  10^{-5}~\mathrm{eV}^2 $$
$$ \tan^{2}\theta=0.40^{+0.10}_{-0.07}.$$
\par This is another proof of the interpretation of solar neutrino deficit
in terms of oscillation.
\par After the results of these two experiments, neutrino oscillations
from a possible theory
become  a well defined physical phenomenon.
\subsection  {Experimental results . Atmospheric  neutrinos }
\par While the observation of solar neutrinos concerns the disappearance
of
$\nue$ and so  the (1,2) mixing parameters, information on the (2,3)
mixing comes from the study of atmospheric  $\numu$ disappearance experiments.
Neutrinos are generated by the decay of hadrons produced by primary
cosmic rays in the upper part of the  atmosphere. $\numu$ are produced by
the decay of pions and kaons while the $\nue$ are produced together with
$\numu$ by the decay of muons.
Many  underground experiments have been made, the one that has given
a definite proof of disappearance of $\numu$ is the Superkamiokande
experiment that  used the same detector used for solar neutrinos.
While the $\nue$  behave according to Montecarlo computations a clear
deficit of upward $\numu$ ie from $\numu$ that have traversed the
Earth was observed . The experiment  studied the double ratio
 R= $(\mu/ e)_{data}/( \mu/ e)_{MC}$ that should be 1 in the absence of
oscillations and that turned out to be  \cite{Skat1}
$$ R=0.658 \pm 0.016 \pm 0.035$$

\par Indications of the upward $\numu$ deficit have been obtained also in 
other
underground detectors \cite{Macro} and \cite{sou}
\par A direct proof of disappearance of muon neutrinos has been
obtained in two long baseline experiments K2K \cite{k2k} and Minos
 \cite {minos}
\par These experiments utilize a two detector scheme. The  distance between
the two detectors
 and the energy of the beams are choosen to access the $\dmq$ region
given by the SK result.  The results of the three  experiments,
either with atmospheric or with accelerators, given
in the following table, are in good agreement
\begin{table}[htbp]
\begin{center}
\begin{tabular}{|c|c|c|}\hline
{\bf experiment} & $\dmq \cdot 10^{-3}~\mathrm{eV}^2$ &
$\sin^{2}2\theta$
\\\hline
ATMO SK~\cite{Skat1}&1.5-3.4& $\ge$ 0.92    \\\hline
K2K~\cite{k2k}~&1.5-3.9  & $\ge$ 0.58 \\\hline
MINOS~\cite{minos}&2.38 $\pm$ 0.2     &$\ge$ 0.98\\\hline
\end{tabular}
\label{dm2}
\caption{limits on the 23 mixing parameters}
\end{center}
\end{table}
 and this fact
is again  a proof of the interpretation of results in terms  of oscillations.
All these experiments are disappearance experiments , the possibility of
$\numu \rightarrow \nue$ is excluded by the SuperK   result in which $\nue$
are
in good agreement with the expectations and by the result  of the Chooz
 reactor
experiment so that the  only possibility left is the $\numu \rightarrow \nutau$ but
no direct evidence of this reaction has been until now given. The Opera
experiment \cite{opera} from CERN to LGNS will look for the appearence of
 $\tau$ produced by $\nutau$ interactions.
The difficulty, given the short lifetime of $\tau$, of detecting  $\nutau$
will be addressed using the high granulariy of  nuclear emulsions.
\subsection{The contribution of Bruno Pontecorvo}
\par As it has been shown in the section \ref{det}
Pontecorvo  suggested  the possibility of measuring the reaction
 $$\nu +Cl^{37} \rightarrow Ar^{37} +e$$
 and as source  of neutrinos he suggested solar neutrinos.
The solar neutrino Davis experiment  started the 'neutrino puzzle'
and the originality of the BP contribution was largely acknowledged.
\par In 1957 R.Davis was doing the same  experiment
 at the Savannah river reactor using $\nueb$.
 a rumor reached B.P in Dubna, that the process had been observed.
The rumour was false and  Davis obtained a null result thus showing a
difference
between neutrinos and antineutrinos.
\par The process
 $$\nueb +Cl^{37} \rightarrow Ar^{37} +e$$

 does not conserve leptonic number, so he  started to think
 to processes that violate
leptonic number and that the reason  that  the reaction  did happen  was
 the transition $\nue \rightarrow
\nueb$ in vacuum.
 He  published  two fundamental papers in 1957:
\par 1)'Mesonium and anti-mesonium', The conclusion of the paper was
 \cite{Pontecorvo:1957cp}
{\sl.. if the conservation of neutrino  charge took no
place the neutrino
antineutrino transition would be in principle possible}.
\par 2)'Inverse beta processes and non conservation of lepton charge'
\cite{Pontecorvo:1957qd}
\par {\sl In the hypothesis of non conservation of neutrino charge

 a beam of neutral leptons from a reactor which at first consists mainly of
antineutrino will change its composition at a certain distance from the reactor}
\par We note that this effect has been observed in the
Kamland experiment.
\par  At that time only one type of neutrinos
was known. After the
discovery
of two types of neutrinos
BP   extended this concept
to flavour oscillations \cite{bile}.
\par In 1967 BP published a paper 'neutrino experiments and the question of lepton charge
conservation' \cite{conse} in which he, given the evidence for $\nu$ and $\numub$
diffference,
 gave examples of processes that could test the leptonic charge conservation.
\par BP was very interested on  the oscillation problem, many times
in collaboration with S. Bilenky. He published several papers on solar neutrinos
\cite{solar1},\cite{astro5}
\cite{solar2},\cite{solar3}. In the last paper he explained  the result of
the Davis experiment in terms
of oscillations.He wrote:
\par
{\sl It appears that the explanation in terms of neutrino mixing is much more
attractive and
natural than other explanations}

\subsection {Present knowledge of the Pontecorvo-Maki-Nakagawa-Sakata
matrix }
\par  As shown in  the chapter on the neutrino oscillation theory
the   mixing matrix U contains 3 angles and possibly  a CP violating term
$\delta$ .The angles $\theta_{12}$ and  $\theta_{23}$ are reasonably well
known while for  $\theta_{13}$ only upper limits are given, the more
stringent has been given by the Chooz Experiment \cite{apollonio};
the phase  $\delta$ is unknown. The determination of  $\theta_{13}$
is very important because a not too small value of this quantity will open
the possibility of the phase  $\delta$  of CP violation in the  neutrino
field to be measured .Experiments have been proposed and are in
preparation ,  T2K
\cite{t2k}
will look in a neutrino beam  from Jaeri Japan to Superkamiokande
through the detection of the subdominat $\numu \rightarrow \nue$ reaction.
The reactor experiment DayaBay \cite{xinheng} will try to improve  the
limit on $\nue \rightarrow \nux$ of
Chooz by  a factor 10

\newpage
\par\LARGE{Appendix}
            
\par \large{ Basics  of neutrino oscillation theory}
\par As in the quark  sector  the weak interactions states are  a linear
superposition of the mass eigenstates.
These states are connected by an unitary matrix U
$$ \nu_{\alpha}=\sum_j U_{\alpha j}\cdot\nu_j $$
with index $\alpha$ running over the three flavor eigenstates and index j
running over the three mass eigenstates.
\par The matrix U is called the Pontecorvo--Maki--Nakagawa--Sakata.
 In the general case, a $3\times3$ matrix can be parametrized by 3 mixing
angles $\theta_1=\theta_{12}$, $\theta_2=\theta_{23}$, $\theta_3=\theta_{13}$
and a CP violating phase $\delta$.
A frequently used parametrization of the U matrix is the following
\begin{eqnarray}\nonumber
U&=&\begin{pmatrix}1&0&0\cr 0&c_{23}&s_{23}\cr 0&-s_{23}&c_{23}
\end{pmatrix}
\begin{pmatrix}c_{13}&0&s_{13}e^{-i\delta}&
 \cr
0&1&0\cr -s_{13}e^{+i\delta}
 &0&c_{13}
\end{pmatrix}
\begin{pmatrix}c_{12}&s_{12}&0\cr -s_{12}&c_{12}&0\cr 0&0&1

\end{pmatrix}
\end{eqnarray}
\par where $c_{jk}=\cos(\theta_{jk})$ ~~~and ~~~~  $s_{jk}=\sin(\theta_{jk})$.
\par  Given three neutrino
masses, we can define two independent
 square mass differences $\dmq_{12}$ and  $\dmq_{23}$.
 $\dmq_{12} =M_{1}^2- M_{2}^2$,
 $\dmq_{23} =M_{2}^2- M_{3}^2$

\par It has  experimentally been shown that
$|\dmq_{12}|\ll |\dmq_{23}|$ and so $\dmq_{13}\simeq
\dmq_{23}$.
\par The mass spectrum is formed by a  closely spaced doublet  $\nu_1$ and
$\nu_2$,
 and by a third state $\nu_3$ relatively distant.
\par In many cases oscillations can be studied considering the simplified
approximation of  two  family
mixing. With two mass states  $M_1$ and $M_2$, the mixing matrix is
reduced to $2\times2$ and is
characterized by a single parameter,the mixing angle $\theta$ (omitting
irrelevant phase factors):

\begin{gather*}
\begin{pmatrix} \cos\theta&\sin\theta\\-\sin\theta&\cos\theta
\end{pmatrix}
\quad
\end{gather*}
\par Consider for example a $\nue$ beam and $\nue ~ \rightarrow \numu$
oscillations
\par At a distance L from
from the source the probability of detecting a $\nue$ as
$\numu$, is
\par $$ P(\nue \rightarrow \numu)
=\sin^{2}(2\theta) \sin^{2}(\Delta
M^{2}L/4E)$$
\par where $\dmq= M_1^{2}-M_2^{2}$.
~Choosing to express $\dmq$ in $eV^2$, L in m(Km) E in MeV(Gev) we have
\par $$ P(\nue \rightarrow \numu)=
\sin^{2}(2\theta) \sin^{2}(1.27 \Delta M^{2}L/E)$$
\par we emphasize that oscillations are sensitive only to the difference
of the square masses.

\bibliographystyle{unsrt}
\section{References}
\bibliography{brunopaper.bbl}
\end{document}